\documentclass{article}

\PassOptionsToPackage{numbers, compress}{natbib}

\usepackage[preprint]{neurips_2024}

\usepackage[utf8]{inputenc} % allow utf-8 input
\usepackage[T1]{fontenc}    % use 8-bit T1 fonts
\usepackage{hyperref}       % hyperlinks
\usepackage{url}            % simple URL typesetting
\usepackage{booktabs}       % professional-quality tables
\usepackage{amsfonts}       % blackboard math symbols
\usepackage{nicefrac}       % compact symbols for 1/2, etc.
\usepackage{microtype}      % microtypography
\usepackage{xcolor}         % colors
\usepackage{ragged2e}
\usepackage{graphicx}
\usepackage{tikz}
\usepackage{amsmath}
\usepackage{algorithm}
\usepackage{algpseudocode}
\usepackage{amssymb}
\usepackage{listings}

\title{Scalable Thermodynamic Second-order Optimization}

\author{
  Kaelan Donatella\thanks{Correspondence to: \texttt{kaelan@normalcomputing.ai}}\,\\
  Normal Computing \\
  \And
  Samuel Duffield\\
  Normal Computing
  \AND
  Denis Melanson\\
  Normal Computing\\
  \And
  Maxwell Aifer\\
  Normal Computing\\
  \And
  Phoebe Klett\\
  Normal Computing
    \And
  Rajath Salegame\\
  Normal Computing
    \And
  Zach Belateche\\
  Normal Computing
  \And
  Gavin Crooks\\
  Normal Computing
    \And
  Antonio J. Martinez\\
  Normal Computing
  \And
  Patrick J. Coles\\
  Normal Computing
}

\begin{document}

\maketitle

\begin{abstract}
Many hardware proposals have aimed to accelerate inference in AI workloads. Less attention has been paid to hardware acceleration of training, despite the enormous societal impact of rapid training of AI models. Physics-based computers, such as thermodynamic computers, offer an efficient means to solve key primitives in AI training algorithms. Optimizers that normally would be computationally out-of-reach (e.g., due to expensive matrix inversions) on digital hardware could be unlocked with physics-based hardware. In this work, we propose a scalable algorithm for employing thermodynamic computers to accelerate a popular second-order optimizer called Kronecker-factored approximate curvature (K-FAC). Our asymptotic complexity analysis predicts increasing advantage with our algorithm as $n$, the number of neurons per layer, increases. Numerical experiments show that even under significant quantization noise, the benefits of second-order optimization can be preserved. Finally, we predict substantial speedups for large-scale vision and graph problems based on realistic hardware characteristics.
\end{abstract}

\section{Introduction}

Due to their fast convergence properties~\cite{amari1998natural,zhang2019fast}, second-order training methods hold great promise to train neural networks efficiently. Such methods form a local quadratic approximation to the landscape and update the parameters by optimizing this approximation within some region~\cite{martens2020new}. Thus, these methods should make more progress per iteration than vanilla gradient descent because of their more detailed modeling of the landscape~\cite{amari1998natural}. Beyond neural network training, second-order methods are also highly popular in reinforcement learning~\cite{kakade2001natural,grondman2012survey}.

Despite these advantages, first-order methods like stochastic gradient descent (SGD) or Adam~\cite{kingma2015adam} are typically used in practical settings. This is largely because of the large computational overhead of inverting the relevant curvature matrix (e.g., the Fisher matrix) in second-order methods. Methods that employ block-diagonal approximations to the curvature matrix, such as Kronecker-Factored Approximate Curvature (K-FAC), have enabled large-scale applications of second-order methods, including to modern neural network architectures with billions of parameters. Nevertheless, their computational costs remain substantially higher than those of first-order optimizers, limiting their practical competitiveness.

Our perspective is that optimizer preference is dictated by the underlying computing hardware that is running that optimizer. While SGD may be preferable on digital hardware, second-order methods could be superior if the computational substrate could accelerate the key bottleneck of these methods. In this work, we investigate the potential for thermodynamic computers to address bottlenecks in second-order optimizers. These computers are physics-based devices that utilize a physical system's tendency to relax to thermodynamic equilibrium as an algorithmic primitive~\cite{conte2019thermodynamic,coles2023thermodynamic_published}.

Recent work showed that thermodynamic computers could unlock a linear speedup (linear in the matrix dimension) when running linear algebraic primitives~\cite{aifer2024_TLA, duffield2023thermodynamic} like solving linear systems, inverting matrices, and exponentiating matrices. Because Natural Gradient Descent (NGD) involves solving a linear system (associated with computing the natural gradient from the standard gradient), running NGD on a thermodynamic computer is predicted to have a linear speedup in the number of neural network parameters~\cite{donatella2024thermodynamic}. However, Thermodynamic NGD was restricted to fine-tuning applications because it is impractical to build a billion-dimensional device that would be required to fully train all parameters of a large-scale neural network. Thus, it remains an open question for how to make Thermodynamic NGD scalable and practical for applications beyond fine tuning.

Thermodynamic NGD is nevertheless appealing because its asymptotic complexity is similar to that of a first-order method, and hence if it could be made practical, one could run second-order methods at the computational cost of a first-order method. In this work, we address the practicality of Thermodynamic NGD by considering the K-FAC algorithm. We show how K-FAC, which employs a block-diagonal approximation to the Fisher matrix, can be turned into a thermodynamic algorithm. 

We show that the asymptotic scaling of our Thermodynamic K-FAC algorithm leads to a linear advantage (i.e., the advantage grows linearly with the width $n$ of the neural network) over standard K-FAC for both the runtime cost and memory cost. Crucially, the block-diagonal nature of K-FAC allows for practical implementation on thermodynamic hardware, as the matrices involved have dimension on the order of a thousand rather than a billion. Hence this provides a scalable means to do (approximate) natural gradient descent with thermodynamic hardware.

To account for the finite precision of thermodynamic hardware, we ran numerical experiments investigating the impact of quantization noise on the performance. We found that even under significant quantization noise, the benefits of second-order optimization (e.g., relative to the Adam optimizer) can be preserved. This suggests some robustness to imprecision for our Thermodynamic K-FAC algorithm.

Finally, our simulations of large-scale vision and language problems show evidence that Thermodynamic K-FAC outperforms both standard K-FAC as well as Adam. Moreover, by altering the hyperparameters of the model (e.g., increasing the width of the network), speedups even larger than those we show in our numerics can be unlocked.

\section{Related work}

There is a large body of theoretical research on NGD~\cite{amari1998natural, martens2020new, bottou2018optimization} arguing that it requires fewer iterations than SGD to converge to the same value of the loss in specific settings. K-FAC~\cite{martens2015optimizing} aims to reduce this complexity and invokes a block-wise approximation of the curvature matrix, which may not always hold. While first introduced for multi-layer perceptrons, K-FAC has been applied to more complex architectures, such as recurrent neural networks~\cite{martens2018kronecker} and transformers~\cite{eschenhagen2024kronecker}, where additional approximations have to be made and where the associated computational overhead can vary. While the K-FAC approximation is uncontrolled, there is a large body of empirical evidence showing its faster convergence per step with respect to Adam and its variants~\cite{martens2015optimizing, martens2018kronecker, eschenhagen2024kronecker, ren2019efficient,gargiani2020promise}. It has also been shown that K-FAC extends the critical batch size for a variety of tasks~\cite{zhang2019algorithmic}, decreasing the diminishing returns usually seen by scaling up the batch size in neural network training.

However, because of the per-step overhead of K-FAC, it remains roughly on-par with first-order methods~\cite{eschenhagen2024kronecker} in terms of per-wall clock time performance. In this work we focus on reducing the runtime per step of the K-FAC optimizer, which directly makes it more competitive.

At the core of our approach are thermodynamic algorithms~\cite{aifer2024_TLA} for solving linear systems and inverting matrices. We remark that alternative analog methods for solving these kinds of problems can be found in Refs.~\cite{sun2019solving,
sun2020time}. In addition, alternative approaches to thermodynamic computing have been proposed~\cite{hylton2020thermodynamic,ganesh2017thermodynamic,lipka2024thermodynamic,whitelam2024thermodynamic}, applications beyond linear algebra have explored such as Bayesian inference~\cite{aifer2024_TBI} and quadratic programming~\cite{bartosik2024thermodynamic}, and closely related to thermodynamic computing is probabilistic computing~\cite{aadit2022massively,kaiser2022life} and reversible computing~\cite{frank2020reversible}.

While several approaches have been proposed to accelerate training of AI models using novel hardware, these efforts typically aim at reducing the constant coefficients appearing in the time cost of computation. For example, analog computing devices have been proposed to achieve reduced
time and energy costs of training relative to available digital technology~\cite{kim2017analog,ambrogio2018equivalent,cristiano2018perspective,aguirre2024hardware}. These devices are generally limited to training a neural network that has a specific architecture (corresponding to the
structure of the analog device). 

\begin{figure*}[t]%
\centering
\scalebox{1}{\input{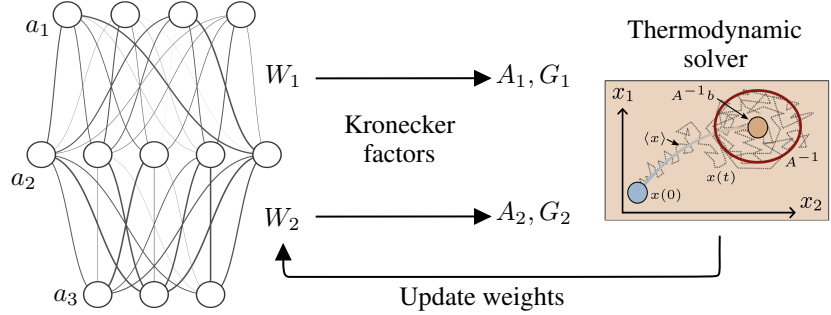}}
\caption{\justifying \textbf{Overview of the thermodynamic algorithm for K-FAC.} On the left is shown a two-layer neural network with weight matrices $W_1$ and $W_2$ and activations $a_1, a_2, a_3$ that are stored on a digital device. From these quantities Kronecker factors $A_\ell$ and $B_\ell$ are computed and sent to the thermodynamic solver, which inverts them or solves a linear system where they enter as the positive semi-definite matrix. Then, the result is sent back to the digital device and the weights are updated. Note that this algorithm is easily parallelized, e.g., many thermodynamic solvers can be used to compute the K-FAC update rule (Eq.~\eqref{eq:K-FAC_update}) for one or more layers each. 
}\label{fig:scheme}
\end{figure*}

A strength of our approach is its flexibility with respect to model architecture. Although this same strength appeared in the Thermodynamic NGD algorithm of Ref.~\cite{donatella2024thermodynamic}, that algorithm would either require (1) a large-scale hardware (with a number of physical components scaling linearly with the number of model parameters) for which important scalability challenges have yet to be solved, or (2) restricting the training tasks only to fine-tuning. In this sense, Ref.~\cite{donatella2024thermodynamic} did not fully solve the issue of training large-scale foundational models. Thus, a key insight of the current paper is to make the training of large-scale AI models practical and scalable for thermodynamic hardware. Moreover, our complexity analysis (Table~\ref{tab:complexities}) suggests that the per-iteration complexity of K-FAC can be made similar to that of a first-order training method.

\section{Natural gradient descent and K-FAC}\label{sec:NGD}

\subsection{Natural gradient descent}
Let us consider a supervised learning setting, where the goal is to minimize an objective function defined as:
\begin{equation}
    \mathcal{L} (\theta) = \frac{1}{|\mathcal{D}|} \sum_{(x,y)\in \mathcal{D}} L(y, f_\theta(x)),
\end{equation}

where $L(y, f_\theta(x))\in \mathbb{R}$ is a loss function and $f_\theta(x)$ is the forward function that is parametrized by $\theta \in \mathbb{R}^N$. These functions depend on input data and labels $(x,y) \in \mathcal{D}$, with $\mathcal{D}$ a given training dataset.

The update rule for Natural Gradient Descent (NGD)~\cite{martens2010deep} is given by
\begin{equation}\label{eq:NGDupdate}
    \theta^{(t+1)} = \theta^{(t)} - \eta \cdot (C^{(t)})^{-1} \nabla_{\theta^{(t)}} \mathcal{L}\left(y, f_{\theta^{(t)}}(x)\right)
%    \theta_{t+1} = \theta_t - \alpha C^{-1}_t \nabla_{\theta_t} \mathcal{L}\left(y, f_{\theta_t}(x)\right)
\end{equation}
where $\eta$ is the learning rate,  $C^{(t)}$ is a pre-conditioning matrix (which can depend on $\theta^{(t)}, x,y$),  and  $\nabla_{\theta^{(t)}} \mathcal{L}\left(y, f_{\theta^{(t)}}(x)\right)$ is the gradient of the objective function. For NGD, the preconditioning matrix is of size $N \times N$, with $N$ the total number of parameters of the model. Therefore, the NGD update quickly becomes impractical for modern neural networks for billions of parameters, as even storing the full pre-conditioning matrix is too expensive.

\subsection{Kronecker-factored approximate curvature (K-FAC)}

%{\color{red}PC: We switch notations from $\theta_t$ to $\theta_{\ell}$. We may want to use one as a subscript and one as a superscript. Later on, we use a double subscript, which is yet another different notation. So we need to standardize our notation. I propose to promote $t$ to a superscript.}

One possible solution to this impractical scaling was proposed in Ref~\cite{martens2015optimizing}, known as Kronecker-factored approximate curvature (K-FAC). This method consists of simplifying the computation of the inverse of the preconditioning matrix $C$ by exploiting its structure.

Let us consider a deep neural network with a layered structure, such that the forward function can be written as:
\begin{equation}
    f_\theta = f_{\theta_L} \circ \dots \circ f_{\theta_2} \circ f_{\theta_1},
\end{equation}
with $\theta = \text{concat}(\theta_1, \theta_2, \dots, \theta_L)$ and $L$ the number of layers;  $\text{concat}(\cdot, \dots, \cdot)$ concatenates vector inputs to a larger vector. For a multilayer perceptron (MLP), we have:
\begin{equation}
\label{eq:MLP-layer}
    f_{\theta_\ell}(a_{\ell-1}) = \phi(W_\ell a_{\ell-1} + v_\ell)= \phi(\overline{W}_\ell \overline{a}_{\ell-1} )
\end{equation}
where $\phi$ is an activation function, $a_{\ell-1} \in \mathbb{R}^{n_{\ell, \mathrm{in}}}$, $W_\ell \in \mathbb{R}^{n_{\ell,\text{out}} \times n_{\ell,\text{in}}}$, and $v_\ell \in \mathbb{R}^{n_{\ell,\text{out}}}$. The bias vector $v_\ell$ can be included as a column in the weight matrix (we denote this expanded weight matrix $\overline{W}_\ell$, and a constant value of unity may be appended to $a_\ell$ (which we denote $\overline{a}_\ell)$, yielding the right-hand side of Eq. \eqref{eq:MLP-layer}.  We also define the pre-activation $s_\ell := \overline{W}_\ell \bar{a}_{\ell - 1}$. The parameter vector can be written as  %\begin{equation}
%    \theta_\ell = \text{concat}(\text{vec}(W_\ell), v_\ell) \in \mathbb{R}^{P_\ell},
%\end{equation}
\begin{equation}
    \theta_\ell = \text{vec}(\overline{W}_\ell),
\end{equation}
where $\text{vec}(\cdot)$  vectorizes a matrix by concatenating its column vectors. The total number of parameters in the $\ell$th layer (i.e., the length of $\theta_{\ell}$) is $P_\ell = n_{\ell,\text{out}} n_{\ell,\text{in}} + n_{\ell,\text{out}}$.

We consider the empirical Fisher as the pre-conditioning matrix and follow Ref.~\cite{martens2015optimizing} (though everything can be extended to the generalised Gauss-Newton matrix with a block-diagonal approximation). In what follows all expectation values are taken over mini-batches of data, with a batchsize $b$. The empirical Fisher is given by:
\begin{equation}
    F := \mathbb{E}[D\theta D\theta^\top]
\end{equation}
where
\begin{equation}
D\theta := [\mathrm{vec}(D\theta_1)^\top, \mathrm{vec}(D\theta_2)^\top, \ldots, \mathrm{vec}(D\theta_L)^\top]^\top, \quad D\theta_{\ell} := \frac{d\mathcal{L}(y, f_\theta(x))}{d\theta_{\ell}}.
\end{equation}
We can therefore rewrite $F$ as:
\begin{align}
    F &= \mathbb{E}\left[[\mathrm{vec}(D\theta_1)^\top, \mathrm{vec}(D\theta_2)^\top, \ldots, \mathrm{vec}(D\theta_L)^\top]^\top[\mathrm{vec}(D\theta_1)^\top, \mathrm{vec}(D\theta_2)^\top, \ldots, \mathrm{vec}(D\theta_L)]\right]\notag\\
     &= \begin{pmatrix}
\mathbb{E}[\mathrm{vec}(D\theta_1)\mathrm{vec}(D\theta_1)^\top]& \mathbb{E}[\mathrm{vec}(D\theta_1)\mathrm{vec}(D\theta_2)^\top]& \ldots & \mathbb{E}[\mathrm{vec}(D\theta_1)\mathrm{vec}(D\theta_L)^\top]\\
\mathbb{E}[\mathrm{vec}(D\theta_2)\mathrm{vec}(D\theta_1)^\top]& \mathbb{E}[\mathrm{vec}(D\theta_2)\mathrm{vec}(D\theta_2)^\top]& \ldots & \mathbb{E}[\mathrm{vec}(D\theta_2)\mathrm{vec}(D\theta_L)^\top]\\
\vdots &\vdots& \ddots & \vdots\\
\mathbb{E}[\mathrm{vec}(D\theta_L)\mathrm{vec}(D\theta_1)^\top]& \mathbb{E}[\mathrm{vec}(D\theta_L)\mathrm{vec}(D\theta_2)^\top]& \ldots & \mathbb{E}[\mathrm{vec}(D\theta_L)\mathrm{vec}(D\theta_L)^\top]
\end{pmatrix}
\end{align}
Thus, we see that $F$ is a block matrix with the $(\ell,\ell')$th block being $F_{\ell,\ell'} = \mathbb{E}[\mathrm{vec}(D\theta_\ell)\mathrm{vec}(D\theta_{\ell'})^\top]$. For an MLP, we have 
\begin{equation}
    D\theta_{\ell} = g_{\ell} \bar{a}_{\ell-1}^\top,\quad\text{where}\quad g_{\ell} := Ds_{\ell}\,.
\end{equation}
Using the identity $\mathrm{vec}(BC^\top) = C \otimes B$, we therefore have the blocks $F_{\ell,\ell'}$ given by
\begin{equation}
F_{\ell,\ell'} = \mathbb{E}[(\bar{a}_{\ell-1}\otimes g_\ell)  ( \bar{a}_{\ell'-1} \otimes g_{\ell'} )^\top] = \mathbb{E}[(\bar{a}_{\ell-1}\otimes g_\ell)  ( \bar{a}^\top_{\ell'-1} \otimes g_{\ell'}^\top )] = \mathbb{E}[(\bar{a}_{\ell-1}\bar{a}^\top_{\ell'-1}\otimes g_{\ell}g_{\ell'}^\top)].
\end{equation}

Crucially, the K-FAC approximation consists in first replacing the average of the Kronecker product by a Kronecker product of averages, as:
\begin{equation}
F_{\ell,\ell'} \approx \mathbb{E}[\bar{a}_{\ell-1}\bar{a}^\top_{\ell'-1}] \otimes \mathbb{E}[ g_{\ell} g_{\ell'}^\top],
\end{equation}
which is an empirically motivated approximation (rather than a theoretically motivated one)~\cite{martens2015optimizing}. We define the following matrices, called the Kronecker factors, as
\begin{equation}\label{eq:A_and_G}
A_{\ell} = \mathbb{E}[\bar{a}_{\ell}\bar{a}^\top_{\ell}]\quad\text{and}\quad G_{\ell} = \mathbb{E}[ g_{\ell}g_{\ell}^\top],
\end{equation}
and we make the further approximation that the Fisher is block-diagonal. Here, the dimensions of $A_{\ell}$ and $G_{\ell}$ are, respectively, $(n_{\ell,\text{in}} +1) \times (n_{\ell,\text{in}}+1)$ and $(n_{\ell,\text{out}}) \times (n_{\ell,\text{out}})$. Note that both $A_{\ell}$ and $G_{\ell}$ are symmetric positive semi-definite (SPSD) matrices.

%{\color{red}PC: Because $A_{\ell}$ and $B_{\ell}$ are so important, I think it would be useful to state their dimensions here. (It wasn't obvious to me what their dimensions were.) Moreover, my understanding is that we are giving them the names "Kronecker factors", but we never explicity say that these are the Kronecker factors, so we should do that. Also, we should state any other properties of these matrices, like whether they are positive semi-definite. I am guessing they are PSD.}

By exploiting the following relations:
\begin{equation}
(B \otimes C)^{-1} = B^{-1} \otimes C^{-1}\quad\text{and}\quad (B \otimes C) \text{vec}(X) =\text{vec}(CXB^\top),
\end{equation}
we can then derive the per-layer parameter update from Eq.~\eqref{eq:NGDupdate}. This results in the K-FAC update rule:
\begin{equation}\label{eq:K-FAC_update}
    \theta_{\ell}^{(t+1)} =  \theta_{\ell}^{(t)} - \alpha u_\ell^{(t)},
\end{equation}
where $u_\ell^{(t)} =  \text{vec}(U_\ell^{(t)})$. The matrix $U_\ell^{(t)}$ is given by 
\begin{equation}\label{eqn:Ult}
    U_\ell^{(t)} = \left(G_\ell^{(t)}\right)^{-1}
    (D \Theta_\ell) 
    \left(A^{(t)}_{\ell - 1}\right)^{-1}
\end{equation}
where $D \Theta_\ell$ is the matrix satisfying $\text{vec}(D \Theta_\ell) = D \theta_\ell$.

%{\color{red}PC: The derivation of the following equation is still hard to follow, so I suggest we explain it in more detail. For example, it looks like we are using the following identity: $(B \otimes C) \text{vec}(X) =\text{vec}(CXB^\top) $. So I suggest that we state this identity in the text.}

%\begin{equation}\label{eq:K-FAC_update}
%    \theta_{\ell}^{(t+1)} =  %\theta_{\ell}^{(t)} - \alpha \cdot \text{vec}[(B_{\ell}^{(t)})^{-1}\text{unvec}(D\theta_\ell) (A_{\ell-1}^{(t)})^{-1}].
%    \theta_{\ell, t+1} =  \tilde{\theta}_{\ell, t} - \eta B_{\ell, t}^{-1}(D\theta_\ell) A_{\ell-1, t}^{-1}.
%\end{equation}

%Let $D \Theta_\ell$ be the matrix satisfying $\text{vec}(D \Theta_\ell) = D \theta_\ell$. Then define $U_\ell^{(t)}$ as
%\begin{equation}
%    U_\ell^{(t)} = \left(G_\ell^{(t)}\right)^{-1}
%    (D \Theta_\ell) 
%    \left(A^{(t)}_{\ell - 1}\right)^{-1}
%\end{equation}
%The parameter update is then
%\begin{equation}\label{eq:K-FAC_update}
%    \theta_{\ell}^{(t+1)} =  \theta_{\ell}^{(t)} - \alpha u_\ell^{(t)},
%\end{equation}
%where $u_\ell^{(t)} =  \text{vec}(U_\ell)$.

%{\color{red}PC: What is $\eta$ here? We need to define $\eta$. If $\eta$ is a learning rate, then note that we previously used $\alpha$ as the learning rate in Eq (2), so we need to standardize our notation.}

Because the Kronecker factors are estimated with minibatches, it is common to compute exponential moving averages (EMA) on them in order to aggregate batch information and reduce minibatch noise. These moving averages, denoted $\tilde{A}_{\ell}^{(t+1)}$ and $\tilde{G}_{\ell}^{(t+1)}$, are defined as:

%{\color{red}PC: Remember to make $t$ a superscript.}

\begin{align}
    \tilde{A}_{\ell}^{(t+1)} &= \beta_A \tilde{A}_{\ell}^{(t)} + (1 - \beta_A) A_{\ell}^{(t+1)} \label{eq:ema_A} \\
     \tilde{G}_{\ell}^{(t+1)} &= \beta_G \tilde{G}_{\ell}^{(t)} + (1 - \beta_G) G_{\ell}^{(t+1)}. \label{eq:ema_B}
    \end{align}

As will be explained further, using an EMA requires to explicitly construct the Kronecker factors, and hence modifies the requirements of the thermodynamic hardware used to accelerate K-FAC.

\subsection{K-FAC for general weight-sharing neural networks}\label{subsec:weight-share}

While the derivation of the K-FAC update shown in Eq.~\eqref{eq:K-FAC_update} is done for MLPs, a similar treatment may be done to any weight-sharing neural network~\cite{eschenhagen2024kronecker}, which includes convolutional neural networks, graph neural networks~\cite{izadi2020optimization}, and transformers~\cite{vaswani2017attention}. This modifies the computation of the Kronecker factors, as the activations $a_\ell$ now are expanded with a weight-sharing dimension of size $R$ (for sequence to sequence models, the sequence length). The gradients $g_\ell$ are also redefined as Jacobians $g_{\ell,r,r'} = \frac{d\mathcal{L}(y, f_\theta(x))_r}{ds_{l,r'}}$. For minibatches of size $b$, the activations are therefore of size $n_{\text{in}}\times b \times R $ (denoted as $a_{{\ell},k,r}$) and the gradients are of size $n_{\text{out}}\times b \times R \times R$ (denoted as $g_{{\ell},k,j, r}$)~\cite{eschenhagen2024kronecker}. The Kronecker factors may therefore be computed as:
\begin{align}
    A_{\ell} &= \frac{1}{bR}\sum_{k, r}^{b, R} a_{{\ell},k,r}a^\top_{{\ell},k,r}\\
    G_{\ell} &= \sum_{k, j,  r}^{b, R, R} g_{{\ell},k,j, r}g^\top_{{\ell},k,j, r}
\end{align}
when the loss has $bR$ terms (in the case of language generation or translation, for example, where tokens have to be compared along the whole sequence when computing the loss). This is the expand setting, and this form of K-FAC is known as K-FAC-expand~\cite{eschenhagen2024kronecker}. For a layer with $n_{\text{in}} = n_{\text{out}} = n$, K-FAC-expand has a per-layer time complexity $\mathcal{O}(bRn^2 + n^3)$ (where the first term comes from calculating the $A_\ell$ and $B_\ell$ factors and the second term from the matrix inverse) and memory $\mathcal{O}(bRn + n^2)$.

When the loss has $b$ terms (in the case of classification), the summation over the weight-sharing dimension may in fact be performed before computing the model output $f_\theta(x)$. This results in gradients being defined as $g_{\ell,r} = \frac{d\mathcal{L}(y, f_\theta(x))}{ds_{l,r}}$ (thus having one less dimension of size $R$). This is known as the reduce setting, in which the Kronecker factors then become:

\begin{align}
    A_{\ell} &= \frac{1}{bR^2}\sum_{k}^{b} \left(\sum_{r}^{R}a_{{\ell},k,r}\right)\left(\sum_{r'}^{R}a^\top_{{\ell},k,r'}\right)\\
    G_{\ell} &= \sum_{k}^{b} \left(\sum_{r}^{R}g_{{\ell},k, r} \right) \left(\sum_{r'}^{R}g^\top_{{\ell},k, r'}\right)
\end{align}

This is known as K-FAC-reduce, and has better computational complexity (due to less sums being performed over the sequence length dimension) while yielding similar results~\cite{eschenhagen2024kronecker} hence is preferred in the reduce setting. K-FAC-reduce has a per-layer time complexity $\mathcal{O}(bn(n+R) + n^3)$ (dominated by computing the $G_\ell$ factors) and memory $\mathcal{O}(bn + n^2)$. 

\section{Accelerating K-FAC with thermodynamic hardware}

\subsection{Potential bottleneck due to matrix inversion}\label{subsec:bottlneck}

%{\color{red}PC:Remember to replace $l$ with $\ell$}

%{\color{red}PC: Remember to make $t$ a superscript.}

Mathematically, K-FAC amounts to estimating the natural gradient for each layer, which involves computing the matrix $U_\ell^{(t)}$ in Eq.~\eqref{eqn:Ult}, repeated here for convenience:
\begin{equation}
    U_\ell^{(t)} = \left(G_\ell^{(t)}\right)^{-1}
    (D \Theta_\ell) 
    \left(A^{(t)}_{\ell - 1}\right)^{-1}\notag
\end{equation}
%\samd{not sure we should repeat (15), or at least not re-number it}
Under certain conditions, the computation of these $U_\ell^{(t)}$ matrices could potentially be a computational bottleneck (for standard digital hardware), due to the matrix inversions required. Specifically, once the Kronecker factors $G_\ell^{(t)}$ and $A^{(t)}_{\ell - 1}$ have been constructed, the computation of $U_\ell^{(t)}$ can be performed with either of the following two methods:
\begin{itemize}
    \item \textbf{Method 1 (Inversion method)} Invert $G_\ell^{(t)}$ and $A^{(t)}_{\ell - 1}, \forall \ell$, then multiply the inverses with $(D \Theta_\ell)$.
    \item \textbf{Method 2 (Linear systems method)} For each column $j$ of the matrix $D\Theta_l$, solve the linear systems $G_\ell^{(t)} x = (D\Theta_l)_j$  to obtain the matrix $Q_\ell^{(t)} = \left(G_\ell^{(t)}\right)^{-1}
    (D \Theta_\ell)$. Then solve the linear systems $(Q_\ell^{(t)})_k = x A^{(t)}_{\ell - 1}$ for each column $k$ of $Q_\ell^{(t)}$.
\end{itemize}
These two methods both have cubic complexities, as for the inversion a Cholesky decomposition or singular value decomposition would be used and in the second case $O(n)$ linear systems are solved with a complexity $O(n^2\kappa)$ if the conjugate gradient method is used, with $\kappa$ the condition number of the matrices involved. For our experiments presented below (in Section~\ref{sec:experiments}), Method~1 was used and only the inversion was profiled and replaced with another solver due to easier implementation. However, we expect Method 2 to be more efficient when using the thermodynamic algorithm presented below as it does not require one to construct the inverse explicitly in memory (and has less operations overall).

%(i) inverting $B_{l,t}$ and $A_{l-1, t}, \forall l$, then multiply the inverses with $(D\theta_l)$ or (ii) solve the linear systems $B_{l, t}x = D\theta_l$ (one for each column of $D\theta_l$) to obtain $B_{l, t}^{-1}(D\theta_l)$, and then solve the linear systems $A_{l-1, t} x = B_{l, t}^{-1}(D\theta_l)$. 

%\subsection{Thermodynamic Algorithm}\label{subsec:thermo_alg}

\subsection{Thermodynamic linear algebra framework}\label{subsec:thermo_alg}

Both of the methods given above for computing $U_\ell^{(t)}$ can be accelerated with thermodynamic hardware~\cite{aifer2024_TLA}. Here we briefly review the thermodynamic linear algebra framework.

Suppose that one is given a positive semi-definite matrix $M$, and the goal is to either to compute the inverse, $M^{-1}$, or to solve a linear system $Mx =b$ for some vector $b$. In either case, one can upload the matrix $M$ to the coupling matrix for a system of coupled harmonic oscillators. This system of coupled harmonic oscillators can, e.g., take the form of RC circuits coupled through capacitive or resistive coupling as discussed in Refs.~\cite{coles2023thermodynamic_published,aifer2024_TLA,melanson2023thermodynamic} (in which case $M$ would be mapped to either the capacitive couplings or the resistive couplings). For more details, see Appendix~\ref{app:hardware}. Moreover, we assume that each oscillator in this system has a stochastic noise source (with inverse temperature $\beta$), and hence the overall system can be viewed as a thermodynamic computer that is called the stochastic processing unit (SPU). 

To solve the relevant linear algebra problem, one allows the SPU to evolve over time according to its natural physical dynamics, which are described by an Ornstein–Uhlenbeck (OU) process. Namely, the dynamics are given by the following stochastic differential equation (SDE):
\begin{equation}
    \label{eq:ODL}
    dx = - (M x - b)  dt +\mathcal{N}\left[0,2\beta^{-1}\, dt\right],
\end{equation}
where the vector $b$ is only relevant to the linear systems case and it can be set to zero if only matrix inversion is desired. (Note that, physically, the $b$ vector corresponds to a local force on each oscillator and hence could corresponds to a locally applied DC voltage in a circuit implementation of an SPU.) 

Allowing the SPU to evolve over time according to Eq.~\eqref{eq:ODL} corresponds to allowing the system to relax to thermodynamic equilibrium. Once it reaches equilibrium, $x$ is Boltzmann distributed, where the Boltzmann distribution is a Gaussian (since the potential is a quadratic form). Specifically, $x$ is distributed according to:
\begin{equation}
    \label{eq:equilibrium-dist}
    x\sim \mathcal{N}[M^{-1}b, \beta^{-1} M^{-1}].
\end{equation}
One can see that the first moment of this distribution is the solution to the linear system $M x = b$, and hence the linear system is solved by sampling $x$ and estimating the mean value. Moreover, second moment of this distribution is proportional to $M^{-1}$, and hence the inverse is computed by estimating the covariances of the samples of $x$.

%$M$ is a positive semi-definite matrix and $\beta$ is a positive scalar (which can be seen as the inverse temperature of the noise). 

\subsection{Thermodynamic K-FAC algorithm}

Thermodynamic linear algebra routines can be directly applied to the K-FAC updates in Eq.~\eqref{eqn:Ult}. In particular, a thermodynamic solver can handle an $n$-dimensional linear system in $\mathcal{O}(n\kappa^2)$ time\footnote{This time complexity emerges from a worst-case upper bound, and there is some evidence that the quadratic dependence on $\kappa$ can be improved in the average-case, which will be published in forthcoming work.}, where $\kappa$ is the matrix condition number. By solving one system per column, we can effectively invert an $n\times n$ matrix in $\mathcal{O}(n^2\kappa^2)$ time, or $\mathcal{O}(n\kappa^2)$ time if the systems are solved in parallel. In the K-FAC setting, this applies directly to the matrix inversions appearing in Method 1, or the linear systems appearing in Method 2 (see Sec.~\ref{subsec:bottlneck} for discussion of these two methods). We also note that the matrix inverses may be directly obtained by estimating the covariance of the stationary distribution given in Eq.~\eqref{eq:equilibrium-dist}. However, Method 2 could be more efficient as it completely avoids matrix-matrix multiplications to compute Eq.~\eqref{eq:K-FAC_update}. Thus the thermodynamic K-FAC algorithm using Method 2 has runtime complexity $\mathcal{O}(n^2\kappa^2)$ as it involves solving $2n$ linear systems.

Furthermore, one may embed the necessary Kronecker factors (e.g., $G_\ell^{(t)}, A_{\ell-1}^{(t)}$) onto thermodynamic hardware for direct, on-device computation of solutions to the corresponding linear systems to reduce the digital memory footprint. This can be done in two ways:
\begin{itemize}
\item Compute and store the Kronecker factors digitally, then transfer them to thermodynamic hardware to perform the linear solves. This is straightforward to parallelize across different thermodynamic solvers, each handling one or more layers.
\item Port activations and gradients directly into hardware, implementing the sums over indices (as in Refs.~\cite{donatella2024thermodynamic} and shown in Appendix~\ref{app:hardware}) on rectangular resistor arrays. This method will be most efficient for small batch sizes and sequence lengths, as the number of physical components scales linearly with these two quantities. For large batches or sequence lengths, it may be more practical to pre-compute the factors rather than porting high-dimensional data to the thermodynamic solver.
\end{itemize}

%The solution to the linear system is found by letting the thermodynamic computer, called the stochastic processing unit (SPU), evolve under an Ornstein–Uhlenbeck (OU) process. The dynamics are given by the following stochastic differential equation (SDE):
%\begin{equation}
%    \label{eq:ODL}
%    dx = - (M x - b)  dt +\mathcal{N}\left[0,2\beta^{-1}\, dt\right],
%\end{equation}
%where $M$ is a positive semi-definite matrix and $\beta$ is a positive scalar (which can be seen as the inverse temperature of the noise). 
%Operationally, one lets the SPU settle to its equilibrium state under the dynamics of Eq.~\eqref{eq:ODL}, at which point $x$ is distributed according to the Boltzmann distribution given by:
%\begin{equation}
%    \label{eq:equilibrium-dist}
%    x\sim \mathcal{N}[M^{-1}b, \beta^{-1} M^{-1}].
%\end{equation}
%One can see that the first moment of this distribution is the solution to the linear system $M x = b$. 
\begin{figure}
    \centering
    \includegraphics[width=\linewidth]{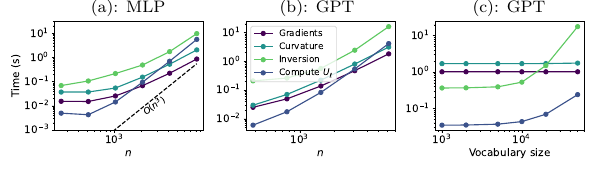}
    \vspace{-0.5cm}\caption{\textbf{Profiling of the K-FAC update for different architectures.} Panel (a): K-FAC update time contributions for an MLP with a fixed depth of 50, with varying number of neurons $n$ on each layer. Panel (b): K-FAC update time contributions for a GPT architecture (based on Ref.~\cite{Karpathy2022}). with varying embedding dimension, which is the number of neurons $n$ in the linear layers. Panel (c): GPT architecture with varying vocabulary size with a fixed embedding dimension. For all plots the reported times are averaged over 10 repetitions and measured on an Nvidia A100 GPU.}
    \label{fig:profiling}
\end{figure}

Finally, note that even if the linear solves or inversions are performed directly on thermodynamic hardware, one still needs explicit Kronecker factors when using an exponential moving average on $G_\ell$ and $A_{\ell-1}$, since one needs to average previously calculated estimates with current values, and this cannot be done by updating activations or gradients directly. Thus, digital computation of these factors remains necessary in this case, leading to a worse runtime and memory complexity when using moving averages.

\subsection{Computational complexity}

% Core K-FAC primitive = matrix inversion + poor scaling
The layerwise block diagonal K-FAC approximation allows the optimizer to scale to very deep neural networks thus avoiding a key limitation of natural gradient descent \cite{amari1998natural, donatella2024thermodynamic}. However, the scaling for wide networks with many neurons per layer remains expensive due to the matrix inversions required in \eqref{eqn:Ult}. Here, the matrices $A$ and $G$ (dropping the layer $\ell$ subscript and iteration $t$ superscript for brevity) have dimension $n \times n$ where $n$ is the number of neurons or width of the layer. Thus the matrix inversion represents the key computational bottleneck for reasonable wide layers, with complexity scaling as $\mathcal{O}(n^3)$ in general implementations. 

% Verify time bottlenecks
In Fig.~\ref{fig:profiling}, the total K-FAC update times is shown for an MLP (panel (a)) and a transformer (GPT)~\cite{Karpathy2022}, for varying widths, in Figs.~\ref{fig:profiling}(a)-(b) and vocabulary sizes, in Fig.~\ref{fig:profiling}(c). For all panels the update time is broken into four components: calculating gradients (through auto-differentiation), constructing the curvature matrices, inverting matrices and computing the final update $U_\ell$ (Eq.\eqref{eq:K-FAC_update}). Throughout this work we consider that the gradient calculation must be performed on a digital device, as it uses auto-differentiation. We observe that as expected in panels (a) and (b), the inversion time and the computation of $U_\ell$ (which contains matrix multiplication) approach cubic scaling as $n$ increases (and is not exactly cubic for low dimensions thanks to parallelization). We also observe that inversion is the main bottleneck across all values of $n$ and for large vocabulary sizes. In the case of large vocabulary sizes, the inversion completely dominates over other contributions, as vocabulary size directly impacts the Kronecker factors of the embedding layers. These measurements are all performed on an Nvidia A100 GPU.

% Thermo speedup of matrix inversion
As described in Section \ref{subsec:thermo_alg}, thermodynamic hardware \cite{coles2023thermodynamic_published} utilizes the physical equilibration of an analog system to efficiently solve linear systems and matrix inversions faster than digital counterparts \cite{aifer2024_TLA}. In both the inversion and linear systems methods described in Section~\ref{subsec:bottlneck}, the thermodynamic solver described in \cite{aifer2024_TLA} costs $\mathcal{O}(n^2\kappa^2)$ for a single layer and matrix condition number $\kappa$. We also note that in practice, the time constant of the hardware's dynamics enters as a constant factor in the runtime, which can be engineered to be extremely small (on the order of a microsecond~\cite{melanson2023thermodynamic, sun2019solving}).

% Discussion Table 1
Table~\ref{tab:complexities} compares the computational single-iteration, single-layer complexities of the introduced thermodynamic K-FAC optimizers (using the linear systems solves to avoid matrix multiplications when computing $U_\ell$) with SGD and digital K-FAC \cite{martens2018kronecker}. SGD and variants such as Adam use a diagonal approximation to the Fisher information and are therefore very cheap to run per iteration, however for practical neural networks, the diagonal approximation to the Fisher is a poor one and leads to many more iterations to reach convergence \cite{martens2015optimizing, anil2020scalable, eschenhagen2024kronecker}. In the simpler case where we do not use the moving average Kronecker factors in Equations~(\ref{eq:ema_A}-\ref{eq:ema_B}), the rectangular components of the matrices in \eqref{eq:A_and_G} can be sent directly to the thermodynamic hardware at a memory cost of $\mathcal{O}(n)$ avoiding the $\mathcal{O}(n^2)$ memory cost of constructing the full Kronecker factors. However, to apply the moving averages (which smooth out the noise from mini-batching), the Kronecker factors need to be constructed. More extensive details on computational complexities for the weight-sharing K-FAC techniques for more general models \cite{eschenhagen2024kronecker} can be found in Table~\ref{tab:full_complexities} in Appendix~\ref{app:commplexites}.

\begin{table}[t!]
%\begin{table}[b]
    \centering
    \renewcommand{\arraystretch}{1.5}
    \begin{tabular}{c|c|c}
        \textbf{Optimizer} & Runtime
        & Memory \\
        \hline
        SGD/Adam &$\mathcal{O}(b n^2)$ & $\mathcal{O}(n^2)$\\
        K-FAC &$\mathcal{O}(bn^2 + n^3)$ & $\mathcal{O}(bn + n^2)$ \\
        Thermodynamic K-FAC  & $\mathcal{O}(bn^2 + n^2\kappa^2)$ & $\mathcal{O}(bn)$  \\
        Thermodynamic K-FAC (w/ EMA) & $\mathcal{O}(bn^2 + n^2\kappa^2)$ & $\mathcal{O}(bn + n^2)$
        \vspace{0.2cm}
    \end{tabular}
    \caption{\textbf{Runtime and memory complexity (per layer) of optimizers considered in this paper.} Here we consider an MLP, where $n$ is the number of neurons per layer, hence there are $n^2$ parameters per layer, and $b$ is the batch size. We also assume that the Kronecker factors all have condition numbers at most $\kappa$. Full complexities that take into account output and weight sharing dimensions for the expand and reduce techniques \citep{eschenhagen2024kronecker} can be found in Table~\ref{tab:full_complexities} in Appendix~\ref{app:commplexites}.
    }
    \label{tab:complexities}
\end{table}

\subsection{Sources of error}\label{sec:quantization}
%\section{Quantization and mitigation of bias}

The Thermodynamic K-FAC algorithm is intended for thermodynamic hardware, an inherently noisy analog platform. The dominant sources of error in this setting include device mismatch, input quantization, and output quantization. Device mismatch arises from fabrication inconsistencies and is thus outside the scope of this paper. Instead, we focus on how quantization errors affect the stability and performance of the optimizer. Quantization noise is general (readout noise with analog-to-digital converters is essentially output quantization) and common to both analog and digital accelerators, making our analysis broadly relevant. In subsequent sections, we strategies to reduce these errors through better quantization schemes. Note that error mitigation methods may also be employed to reduce such errors~\cite{aifer2024error}, and that the thermodynamic nature of our algorithm makes it inherently robust to thermal noise.

%Our proposed thermodynamic algorithm for K-FAC would be run on hardware with limited precision, leading to quantization of both the input and the output.

%When contemplating building an accelerator for the task of solving a linear system (or of matrix inversion), a natural question emerges: what is the effect of errors on the performance and stability of the solver? In this study, we investigated specifically the effect of quantization on the performance of the algorithm. In this context, quantization can take place either on the input or the output of the solver.

\subsubsection{Input quantization}

Given a positive semi-definite input matrix $M$, the thermodynamic hardware stores an approximate quantized version, $\tilde{M}$, up to a certain level of bit-precision in integer format. This quantization can make $\tilde{M}$ no longer positive semi-definite, thus creating instabilities in the thermodynamic system. This would have catastrophic effect on the solver and the algorithm would simply fail. The traditional method to overcome this issue is to ensure that the hardware has more bits of precision for each element of the matrix, such as 32 or 64 bits~\cite{anil2020scalable}, in order to not have the quantization error make the matrix non-definite. This approach works well, but is quite costly in terms of resources. An alternative method to overcome the problem is to modify how we quantize and store the input matrix using a conservative quantizer, that is, a quantizer that conserves the definiteness of the input matrix. In our numerical experiments (presented below), we use a diagonal-dominant quantization method~\cite{funk2021conservative}: 
\begin{itemize}
    \item Round each off-diagonal matrix element to the nearest value available in the hardware.
    \item Calculate the sum of off-diagonal rounding errors in each row of the matrix.
    \item Add the sum of rounding errors to the diagonal.
    \item Round all the shifted diagonal elements up.
\end{itemize}
While this method ensures the definiteness of the quantized matrix, it increases the error significantly in some cases since it adds an error term proportional to the dimension of the matrix. We note that alternative approaches can be used to make the diagonal shift smaller and thus reduce the error, but these are either more computationally demanding~\cite{funk2021conservative} or are based on empirically determined constants that are application-specific~\cite{higham2021precision}.

\subsubsection{Output quantization}
An additional source of quantization error one can expect from a dedicated linear algebra accelerator is the output quantization. This quantization comes from the conversion of the precision of the accelerator and the precision of the rest of the computation. For example, if the accelerator is analog, then the output quantization comes from the analog to digital conversion.    

%\subsubsection{Randomized quantization}\label{sec:quantization}

\section{Experiments}\label{sec:experiments}

\subsection{AlgoPerf experiments}

\begin{figure}
    \centering
    \includegraphics[width=\linewidth]{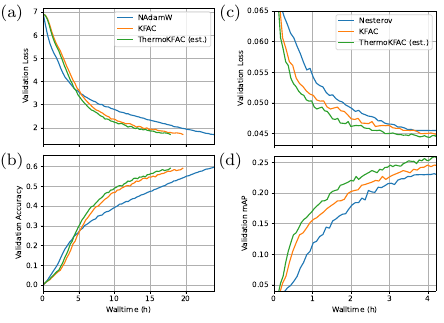}
    \caption{\textbf{Results on ImageNet and OGBG.} Panels (a-b): validation loss and validation accuracy for the NAdamW (the baseline given by AlgoPerf), K-FAC and Thermodynamic K-FAC (estimated) optimizers as a function of the wall-clock time for training a ViT on ImageNet. Panels (c-d): validation loss and validation mean-average precision (mAP) for the Nesterov (baseline), K-FAC and Thermodynamic K-FAC (estimated) optimizers as a function of the wall-clock time for training a GNN on ogbg-molpcba. For the baselines, the hyperparameters are directly taken from the AlgoPerf benchmark and were tuned for the K-FAC optimizers (see Appendix~\ref{sec:code}).}
    \label{fig:algoperf}
\end{figure}

In this section we present numerical results of simulating the thermodynamic K-FAC method on AlgoPerf workloads~\cite{dahl2023benchmarking}. These workloads provide strong baselines and fixed model architectures, enabling direct comparisons of training algorithms without confounding factors such as architecture changes. We consider two workloads: training a vision transformer (ViT) on ImageNet, and training a graph neural network (GNN) on the ogbg-molpcba data set, a popular property prediction dataset for small molecules. We note that these datasets were also used for benchmarking in Ref.~\cite{eschenhagen2024kronecker}, which introduced K-FAC-reduce and K-FAC-expand.

\subsubsection{ViT on ImageNet}

The first workload that we consider is training a ViT on ImageNet. This task involves a state-of-the-art vision model, on a challenging image classification dataset. The baseline optimizer for this workload is Nesterov-accelerated Adam with weight-decay (NAdamW)~\cite{loshchilov2017decoupled}. For these experiments as with others in this work, the baselines use the hyperparameters from the AlgoPerf paper, thus setting a robust baseline to beat that was found externally after extensive tuning and benchmarking. In Fig.~\ref{fig:algoperf}(a-b), the achieved validation loss and the validation accuracy are shown as a function of wall-clock time, that is measured for the baseline and the K-FAC optimizer, and estimated for Thermodynamic K-FAC. This estimation is done by measuring the fraction of the computation time spent on the inverse, which in this case is $11\%$, and estimating a speedup on the matrix inversion for the dimensions considered based on numerics for the matrix inversion primitive, with assumptions detailed in the Appendix and in related work such as Refs.~\cite{melanson2023thermodynamic, aifer2024_TLA,donatella2024thermodynamic}.

\subsubsection{GNN on ogbg-molpcba}

Another workload we consider is training a GNN on ogbg-molpcba, with our results plotted in Fig.~\ref{fig:algoperf}(c-d). The baseline optimizer for this workload is Nesterov~\cite{sutskever2013importance}. For the K-FAC optimizer, here the fraction of the computation time spent on the inversions is $27\%$, meaning a larger speedup can be unlocked, reaching an overall speedup to reach the same validation metrics than the baseline by about $50\%$.

%\subsubsection{Transformer on WMT}

\begin{figure}
    \centering
    \includegraphics[width=\linewidth]{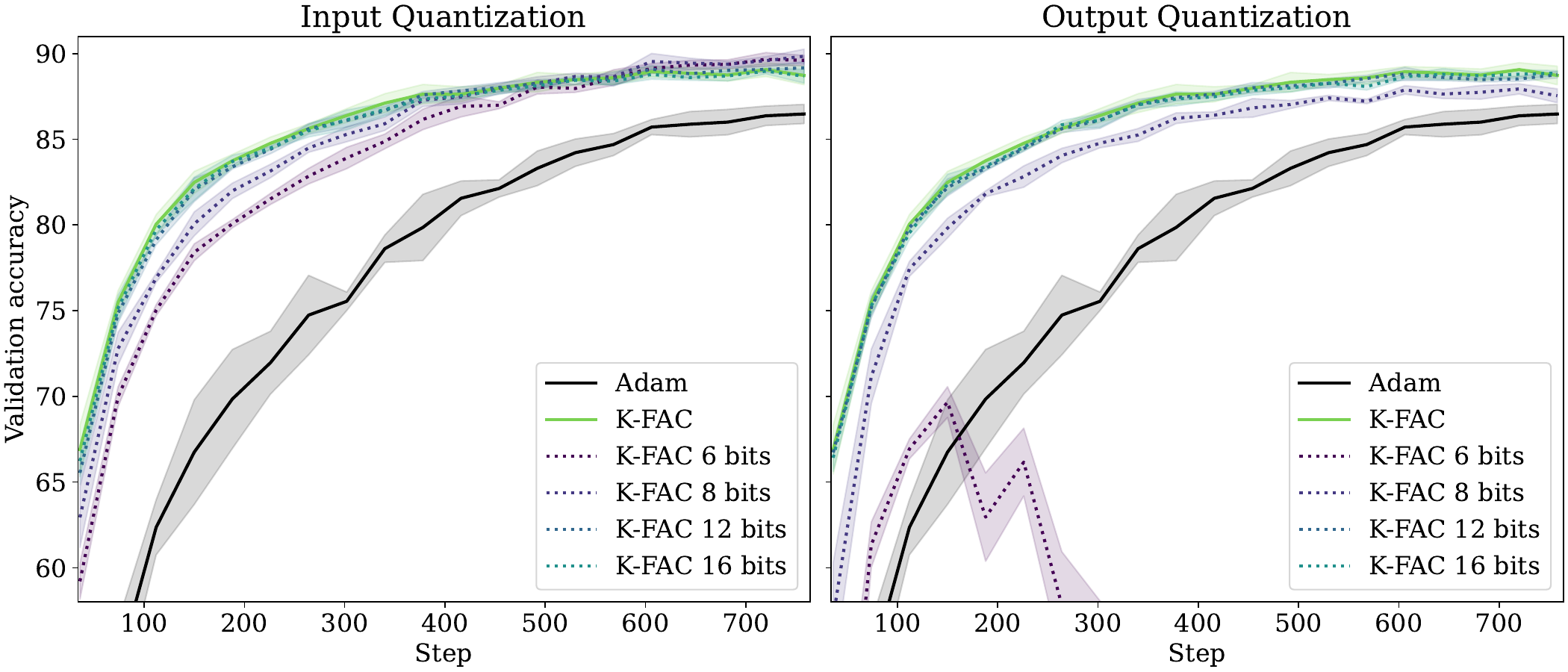}
    \caption{\textbf{Effect of quantization on K-FAC training accuracy.} Validation accuracy from training a ResNet on image classification with either the Adam optimizer or with the K-FAC optimizer for various levels of precision (integer 6, 8, 12 and 16 bits, and floating-point 32 bits at full precision). The left panel shows the effects of input quantization while the right panel shows the effect of output quantization. The lines correspond to the mean values over 5 runs, while the shaded areas represent one standard deviation away from the mean. Brighter colours indicate higher precision.}
    \label{fig:quantization}
\end{figure}

\subsection{Quantization experiments}

To assess the effect of the input and output quantization on a real task, we ran an image classification task on a ResNet using K-FAC with input or output integer quantization. The results are plotted in Fig.~\ref{fig:quantization}. According to these results, the input quantization does not seem to have a large impact on accuracy. We believe that this is partly due to an effective conditioning of the matrix (similar to damping) brought on by the specific diagonally dominant quantization scheme used. The impact of output quantization on training accuracy is more pronounced as compared to input quantization. This provides guidance for how to build a potential hardware architecture for this application, highlighting the need for an output resolution of at least 8 bits while being much less sensitive to input quantization. The broader message of Fig.~\ref{fig:quantization} is a general robustness to quantization that K-FAC appears to have, as even an 8-bit K-FAC optimizer is competitive against a full precision Adam optimizer. This lends hope that a moderate precision hardware architecture could be employed to accelerate K-FAC, with good performance expected.

Moreover, we remark that error mitigation methods have been developed for thermodynamic computing that effectively boost the input precision by several bits~\cite{aifer2024error}. Similar methods could likely be developed for output precision. These error mitigation methods could be incorporated into our Thermodynamic K-FAC optimizer to further enhance performance.

\section{Conclusion}

%--> bringing complexity of 2nd order method down to that of a 1st order method (with respect to $n$)

%--> 
This work introduces a scalable approach to second-order optimization by leveraging thermodynamic hardware to accelerate the K-FAC optimizer. By offloading operations to a physical system that efficiently solves linear algebraic primitives, we achieve a significant reduction in per-iteration asymptotic runtime complexity. This enables K-FAC to approach the efficiency of first-order methods while preserving the convergence benefits of second-order optimization on AlgoPerf benchmarks.

%This leads to asymptotic gains due to the superior runtime complexity of the thermodynamic algorithm for matrix inversion, as well as an improvement in memory requirements for when no moving averages are used.

Experimental results suggest that our method is robust to quantization noise, making it a viable candidate for low-precision analog or mixed-signal implementations. Additionally, our complexity analysis highlights a growing advantage over digital K-FAC as the network width increases.

Extensions of this approach include integrating thermodynamic solvers into large-scale deep learning pipelines and investigating alternative hardware accelerators for second-order methods. Additionally, optimizing the interaction between digital and thermodynamic components could further improve the practicality of our method for real-world training workloads.
Finally, this approach could be extended to various other local approximations to the objective function, which may be accurate over a larger region. This could potentially further reduce the number of optimization iterations necessary over first and second-order methods and remains an important direction for future work.

%Consider the following energy-based viewpoint of our work. Training a machine learning model can be formulated as the task of minimizing an energy function; it is therefore a natural application for thermodynamic computing devices, where the damped stochastic dynamics of a system minimizes its physical free energy.

%Second-order training methods like NGD employ a local quadratic approximation to the loss landscape and then optimize that approximate landscape~\cite{martens2020new}. Given a thermodynamic device with quadratic potential energy (e.g., the coupled harmonic oscillator hardware discussed in our work), the local approximation to the objective function can be mapped to the device's potential energy, allowing the optimization problem to be solved via the physical dissipation of energy. This highlights the deep connection between NGD and dissipative harmonic oscillator systems.

%(e.g., the coupled harmonic oscillator hardware discussed in our work)

%When the objective function is approximately quadratic in some region of the loss landscape, the objective in this region can be mapped to the energy of a thermodynamic device with quadratic potential energy (e.g., the coupled harmonic oscillator hardware proposed discussed in our work). Solving an optimization problem via local quadratic approximation is the basis of second-order training methods like NGD~\cite{martens2020new}.

%Non-gaussian

%Second-order training methods like NGD employ a local quadratic approximation to the landscape than then optimize that approximate landscape~\cite{martens2020new}.

\section{Acknowledgements}

Thanks to the Advanced Research + Invention Agency's (ARIA) Scaling Compute programme for funding this work.

%We acknowledge support from the Advanced Research Invention Agency (ARIA), as part of the Scaling Compute program.

\bibliographystyle{unsrtnat}
%\bibliography{tbi.bib}
%\bibliography{thermo.bib}
\bibliography{ThermoMasterBib.bib}

\appendix

\section{Energy-based viewpoint}

Consider the following energy-based viewpoint of our work. Training a machine learning model can be formulated as the task of minimizing an energy function; it is therefore a natural application for thermodynamic computing devices, where the damped stochastic dynamics of a system minimizes its physical free energy.

Second-order training methods like NGD employ a local quadratic approximation to the loss landscape and then optimize that approximate landscape~\cite{martens2020new}. Given a thermodynamic device with quadratic potential energy (e.g., the coupled harmonic oscillator hardware discussed in our work), the local approximation to the objective function can be mapped to the device's potential energy, allowing the optimization problem to be solved via the physical dissipation of energy. This highlights the deep connection between NGD and dissipative harmonic oscillator systems.

Assuming a thermodynamic device with a non-quadratic potential energy, this approach could be extended to various other local approximations to the objective function, which may be accurate over a larger region. This could potentially further reduce the number of optimization iterations necessary over first and second-order methods, and remains an important direction for future work.

\section{Computational Complexities}\label{app:commplexites}

Table~\ref{tab:full_complexities} expands on Table~\ref{tab:complexities} for the reduce and expand versions of K-FAC \cite{eschenhagen2024kronecker}. These versions of K-FAC extend to more general weight-sharing architectures such as CNNs and transformers and when using the generalised Gauss-Newton matrix as the curvature matrix. The complexities are more subtle due to the additional complexity added to extend to general weight-sharing architectures - in particular added dependence on the output and weight-sharing dimensions. However the cubic to quadratic runtime speedup via the thermodynamic hardware in terms of the width of the network (i.e., the number of neurons per layer) remains.

\begin{table}[h!]
%\begin{table}[b]
    \centering
    \renewcommand{\arraystretch}{1.5}
    \begin{tabular}{c|c|c}
        \textbf{Optimizer} & Runtime
        & Memory \\
        \hline
        SGD/Adam &$\mathcal{O}(bn^2)$ & $\mathcal{O}(n^2)$\\
        K-FAC-reduce &$\mathcal{O}(bCn(C+n+R) + n^3)$ & $\mathcal{O}(bn + n^2)$ \\
       
        Thermodynamic K-FAC-reduce  & $\mathcal{O}(bCn(C+n+R) + n^2\kappa^2)$ & $\mathcal{O}(bn)$  \\
        Thermodynamic K-FAC-reduce (w/ EMA) & $\mathcal{O}(bCn(C+n+R) + n^2\kappa^2)$ & $\mathcal{O}(bn + n^2)$  \\
         K-FAC-expand &$\mathcal{O}( bRCn(C+n) + n^3)$& $\mathcal{O}(bRn + n^2)$ \\
        Thermodynamic K-FAC-expand & $\mathcal{O}(bRCn(C+n) + n^2\kappa^2)$ & $\mathcal{O}(bRn)$ \\
        Thermodynamic K-FAC-expand (w/ EMA) & $\mathcal{O}(bRCn(C+n) + n^2\kappa^2)$ & $\mathcal{O}(bRn + n^2)$ 
        \vspace{0.2cm}
    \end{tabular}
    \caption{\textbf{Runtime and memory complexity (per layer) of optimizers considered in this paper.} Here $n$ is the number of neurons per layer, hence there are $n^2$ parameters per layer, and $b$ is the batch size. $C$ is the output dimension, and $R$ is the weight-sharing dimension. We assume that the Kronecker factors have condition number at most $\kappa$. Expand and reduce refer to the weight-sharing techniques described in Section~\ref{subsec:weight-share} and Ref.~\cite{eschenhagen2024kronecker}.}
    \label{tab:full_complexities}
\end{table}

\section{Code implementation and hyperparameters}\label{sec:code}

Our code implementation is based on \texttt{asdl}~\cite{osawa2023asdl} with an addition of K-FAC-reduce and K-FAC-expand~\cite{eschenhagen2024kronecker} for linear and convolutional layers. 
For the experiments reported in Fig~\ref{fig:profiling}(a), an MLP with $50$ layers was used, with inputs being CIFAR-10 images and a batch size $b=512$. For the GPT in Fig~\ref{fig:profiling}(b), the vocabulary size is fixed to $10^4$, with a sequence length $R=64$, a batch size $b=64$ and 3 transformer layers. (this was necessary for the model to fit in memory on a single GPU). The GPT results in Fig~\ref{fig:profiling}(c) were obtained with $b=64, R =512$, 12 layers and $n=768$.

For the K-FAC training experiments, Table~\ref{tab:hyperparams} shows the relevant hyperparameters for the experiments we performed. All were performed with the conservative quantization method explained in Section~\ref{sec:quantization}.

\begin{table}[h!]
%\begin{table}[b]
    \centering
    \renewcommand{\arraystretch}{1.5}
    \begin{tabular}{c|c|c|c|c|c|c}
        Experiment & Inv. I/O resolution
        & Machine & Optimizer & LR & Damping & EMA decay \\
        \hline
        CIFAR-10 & (varying) & 1xA100 
& SGD & 0.1 &$0.001$ & $0.9999$ \\
        ImageNet-ViT & 12/12 &4xV100 & NAdamW& 0.0012 & $0.001$& $0.999$  \\
       
        OGBG-GNN & 12/12 & 4xV100 &Nesterov & 10 &  $0.005$ &  $0.999$ 
         \vspace{0.2cm}
    \end{tabular}
    \caption{\textbf{Hyperparameters and configurations for experiments.} For the AlgoPerf experiments, we combined K-FAC with other optimizers as in Ref.~\cite{eschenhagen2024kronecker} as it showed superior performance over the configurations we tested. We leave the exploration of why this leads to superior performance to future work.}
    \label{tab:hyperparams}
\end{table}

\section{Inversion time for AlgoPerf workloads in a multi-GPU setting}

We profiled the time spent on matrix inversion in multi-GPU environments, where additional synchronization and communication overhead reduces the fraction of total compute time devoted to inversion. To validate this, we measured the same workloads on an NVIDIA 4×V100 system and on an NVIDIA 8×A100 system with NVLink (providing better communication bandwidth). Table~\ref{tab:profiling} shows that higher communication bandwidth shifts the bottleneck back to compute, increasing the fraction of time spent on inversion. We also note that the current \texttt{asdl} K-FAC implementation is not heavily optimized for further offloading of inversions; thus, inversion cost may still not dominate overall runtime. Similarly, the linear-systems approach (Method 2) described in the main text should increase this computational load.

\begin{table}[h!]
%\begin{table}[b]
    \centering
    \renewcommand{\arraystretch}{1.5}
    \begin{tabular}{c|c|c}
        Workload & 4xV100
        & 8xA100 \\
        \hline
        ImageNet-ViT & $11\%$ &  $16\%$\\
       
        OGBG-GNN   &$27\%$ & $35\%$  
         \vspace{0.2cm}
    \end{tabular}
    \caption{\textbf{Total time spent on inversion for AlgoPerf workloads.} Shown are the percentages for 4×V100 and 8×A100 systems. The 8×A100’s improved communication bandwidth reduces overhead elsewhere, increasing the share of total runtime spent on inversion.}
    \label{tab:profiling}
\end{table}

\section{Potential hardware architecture}\label{app:hardware}

The thermodynamic K-FAC algorithm can be implemented via a similar hardware architecture to what is presented in Refs.~\cite{aifer2024_TLA, melanson2023thermodynamic,donatella2024thermodynamic}. The Kronecker factors can be digitally constructed and sent onto a system whose evolution is described by the differential equation (we take here $A_\ell$ but it is equally valid for the $G_\ell$'s):
\begin{equation}\label{eq:app1}
dV = -( A_\ell+ \lambda \mathbb{I}) Vdt- bdt + \mathcal{N}(0, 2\kappa_0dt)
\end{equation}where $\kappa_0$ is the noise variance and $V = (V_1, V_2, \ldots, V_N)$ is a vector of voltages.

\begin{figure}
    \centering
    \includegraphics[width=0.8\linewidth]{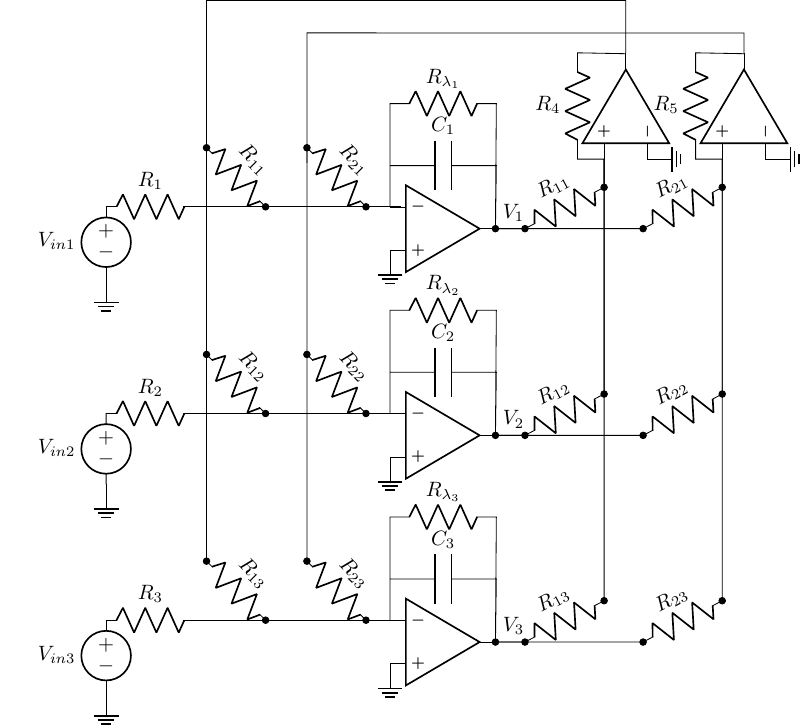}
    \caption{\textbf{Circuit diagram of a possible implementation of the thermodynamic solver.}}
    \label{fig:circuit}
\end{figure}

One may only consider the activations $a_\ell$ and gradients $g_\ell$ that enter the construction of $A_\ell$ and $B_\ell$ respectively and send them directly onto similar hardware. This alternative implementation is comprised of two arrays of resistors of size $(b, N), (b, N)$ for encoding $a_\ell$ and $a_\ell^\top$, respectively. These arrays of resistors enable one to implement the following differential equation in hardware:
\begin{equation}\label{eq:app2}
dV = -( a_\ell a_\ell^\top+ \lambda \mathbb{I}) Vdt- bdt + \mathcal{N}(0, 2\kappa_0dt).
\end{equation}
This system may be implemented with the circuit diagram shown in Fig.~\ref{fig:circuit}, where $N = 3$, $b=2$. We assume the capacitors all have the same value $C$, and the resistors with no labels all have the same value $R_0$. By Kirchhoff’s current law, we obtain the equation of motion for the voltage vector $V = (V_1, V_2, V_3)$ as:

$$C\dot{V} = -(\mathcal{G}V + \lambda V - R^{-1}V_{in})
$$
with $V_{in} = (V_{in1}, V_{in2}, V_{in3})$, $R = \mathrm{diag}(R_1, R_2, R_3)$, $\lambda = \mathrm{diag}(1/R_{\lambda_1}, 1/R_{\lambda_2}, 1/R_{\lambda_3})$. We have

\begin{equation}
    \mathcal{G} =  a_\ell a_\ell^\top = \begin{pmatrix}\frac{1}{R^a_{11}} & \frac{1}{R^a_{12}} \\\frac{1}{R^a_{21}} & \frac{1}{R^a_{22}} \\\frac{1}{R^a_{31}} & \frac{1}{R^a_{32}} \end{pmatrix} 
\begin{pmatrix}\frac{1}{R^a_{11}} & \frac{1}{R^a_{21}} &\frac{1}{R^a_{31}}\\\frac{1}{R^a_{12}} & \frac{1}{R^a_{22}} & \frac{1}{R^a_{32}}\end{pmatrix} \frac{1}{R_0^2},
\end{equation} where we therefore have a set of resistors $R^a$ representing the $a$ tensor and its transpose. 
At steady state the average voltage vector corresponds to the natural gradient estimate, since  for $\dot{V} = 0$, the average voltage vector is $\langle V\rangle = \mathcal{G}^{-1}R^{-1}V_\mathrm{in}$, which corresponds to the solution of the linear system $Ax = b$ with $A = \mathcal{G}$, $x = V$, $b = R^{-1}V_\mathrm{in}$. 
\\The resistor values $R^{a}_{ij}$ can directly be calculated as $1/a_{ij}$ (or $1/a_{ji}$ for the transpose), and the total number of resistors in the circuit is $2bN$ (12 in the schematic shown). One may operate the thermodynamic linear solver by setting the voltage values $V_\mathrm{in}$ to the rows of the gradient matrix $(D \Theta_\ell)$ with a digital-to-analog converter, and set the values of the programmable resistors thanks to a digital controller. The time for the system to relax to equlibrium (and therefore for the linear system to be solved) is:

$$ \tau = \frac{RC}{\alpha_{\min}} $$

where $R$ is a resistance scale (which means that all resistances  $R_{ij}$  are a multiple of this), $C$ is the capacitance (assuming all the capacitances are the same), and  $\alpha_{\min}$  is the smallest eigenvalue of the (unitless) $\mathcal{G}$  matrix. After this time, all the modes of the system will have relaxed, which may be too conservative (for example, in the case where there is only one slow mode, and all other modes are fast). With regularization,  $\alpha_{\min}$ is lower-bounded by the regularization factor  $\lambda$  (which is between $10^{-3}$ and $0.5$ for all experiments). For timing purposes, $RC$ is kept as the relaxation time, because of the problem-dependence of  $\alpha_{\min}$. The estimated speedups on the matrix inverse primitive (see also Refs.~\cite{aifer2024_TLA, melanson2023thermodynamic}) are based on the following assumptions:
\begin{itemize}
    \item 16 bits of precision (less bits of precision will lead to weaker encoding  requirements, therefore a larger speedup).
    \item A digital transfer speed of $ 50$ Gb/s.
    \item $R = 10^3 \,\Omega$, $C= 1 \,\text{nF}$, which means $RC = 1 \mu \text{s}$ is the characteristic timescale of the system.
\end{itemize}

\end{document}